\documentclass[a4paper,11pt,onecolumn]{article}%
\usepackage{amsmath}
\usepackage{amsfonts}
\usepackage{amssymb}
\usepackage{graphicx}%
\providecommand{\U}[1]{\protect\rule{.1in}{.1in}}
\providecommand{\U}[1]{\protect\rule{.1in}{.1in}}
\providecommand{\U}[1]{\protect\rule{.1in}{.1in}}
\providecommand{\U}[1]{\protect\rule{.1in}{.1in}}
\providecommand{\U}[1]{\protect\rule{.1in}{.1in}}
\providecommand{\U}[1]{\protect\rule{.1in}{.1in}}
\providecommand{\U}[1]{\protect\rule{.1in}{.1in}}
\providecommand{\U}[1]{\protect\rule{.1in}{.1in}}
\providecommand{\U}[1]{\protect\rule{.1in}{.1in}}
\providecommand{\U}[1]{\protect\rule{.1in}{.1in}}
\providecommand{\U}[1]{\protect\rule{.1in}{.1in}}
\providecommand{\U}[1]{\protect\rule{.1in}{.1in}}
\providecommand{\U}[1]{\protect\rule{.1in}{.1in}}
\providecommand{\U}[1]{\protect\rule{.1in}{.1in}}
\providecommand{\U}[1]{\protect\rule{.1in}{.1in}}
\providecommand{\U}[1]{\protect\rule{.1in}{.1in}}
\providecommand{\U}[1]{\protect\rule{.1in}{.1in}}

\begin{document}

\title{Modified Sch\"{o}dinger dynamics with attractive densities}
\author{F. Lalo\"{e}\\LKB, ENS and CNRS, 24 rue Lhomond, 75005\ Paris, France}
\date{\today}
\maketitle

\begin{abstract}
The linear Schr\"{o}dinger equation does not predict that macroscopic bodies
should be located at one place only, or that the outcome of a measurement
shoud be unique. Quantum mechanics textbooks generally solve the problem by
introducing the projection postulate, which forces definite values to emerge
during measurements; many other interpretations have also been
proposed.\ Here, in the same spirit as the GRW and CSL theories, we modify the
Schr\"{o}dinger equation in a way that efficiently cancels macroscopic density
fluctuations in space.\ Nevertheless, we do not assume a stochastic dynamics
as in GRW or CSL theories. Instead, we propose a deterministic evolution that
includes an attraction term towards the averaged density in space of the de
Broglie-Bohm position of particles, and show that this is sufficient to ensure
macroscopic uniqueness and compatibility with the Born rule. The state vector
can then be seen as directly related to physical reality.

\end{abstract}

\begin{center}
**********
\end{center}

Macroscopic uniqueness is not a natural physical consequence of standard
quantum mechanics. This is because the linear Schr\"{o}dinger equation can
lead to situations where the position of macroscopic physical systems (the
pointer of a measurement apparatus for instance) have non-zero probabilities
to be at the same time at very different points of space. This difficulty is
illustrated by the famous Schr\"{o}dinger cat thought experiment:\ the linear
evolution of the state vector leads to a state containing at the same time a
dead and an alive cat.\ Schr\"{o}dinger considers these superpositions of
completely different macroscopic states as a \textquotedblleft quite
ridiculous case\textquotedblright\ \cite{Schrodinger-cat, Trimmer,
Wheeler-Zurek}.\ The problem is that nothing in the dynamical equations can
reduce the big fluctuations of the macroscopic density of particles that then
occur.\ But macroscopic physical objects occupying simultaneously completely
distinct positions have never been observed; when experiments are performed, a
single position of the macroscopic measurement pointer seem to appear for each realization.

Many interpretations of quantum mechanics have been proposed to deal with this
apparent contradiction.\ Historically, the projection postulate was introduced
(but not approved by Bohr) by Heisenberg \cite{Heisenberg-1927} and von
Neumann \cite{von-Neumann-1932}, who started from an analysis of the
measurement process in terms of quantum mechanics\cite{Jammer}. Von Neumann
uses the Schr\"{o}dinger equation to study the behavior of a chain of
measurement apparatuses, and finds that no definite result will ever be
obtained, even after a long chain of measurements -- this difficulty is known
as the \textquotedblleft infinite von Neumann regress\textquotedblright.\ He
solves it by introducing his projection postulate, which assumes that a sudden
change of the of the state vector is introduced, in order to update it with
the information gained in a measurement.\ At the other extreme, in the Everett
interpretation \cite{Everett}, the problem is solved by considering
macroscopic uniqueness is not a physical phenomenon, but a delusion arising
from the very functioning of the memory registers of human minds. Numerous
other interpretations have been proposed \cite{Laloe}: modal, relational,
consistent histories, informational, etc.\ Most of them do not change the
standard equations of quantum mechanics, but focus on the best way to
interpret the state vector, its relation with physical reality, information,
experimental context, etc.\ All these interpretations are interesting, but for
the moment none has emerged as the universally accepted optimal point of view.

Other families of interpretations consider that the problem of reconciling
quantum dynamics with what seems to be a routine observation, namely the
uniqueness of the classical world, should be taken seriously: the formalism
and equations of quantum mechanics should be adapted to predict this
uniqueness without ambiguity. The two best known categories are the de
Broglie-Bohm (dBB) interpretation \cite{de-Broglie, Bohm, Holland,
Duerr-et-al} and the spontaneous localization theories, either in the original
GRW discontinuous form \cite{GRW}, or in the continuous CSL form \cite{CSL} --
for a review, see for instance \cite{Bassi-Ghirardi}. In the dBB theory,
particle positions moving in ordinary 3D space are added to the variables of
standard quantum mechanics; these positions have uncontrollable random initial
values, but then move in a perfectly deterministic way. They are considered as
physically real. In the GRW and CSL theories, no additional variables are
assumed, but the usual Hamiltonian in the equation of motion of the state
vector gets additional stochastic terms; the wave function is considered as a
field propagating in configuration space (not ordinary 3D space) under the
effect of fundamentally random processes. In these theories, measurement
processes are not seen as special events, but just ordinary interaction
processes between a measured system and apparatus; the observer is not a
necessary ingredient of the theory.

Here we propose a combination of these two theories, where the dBB positions
are still part of the dynamical equations: the positions are driven by the
wave function (as in the dBB theory), but they also react on it (which does
not occur in the dBB theory). The dynamics of this process is very different
from that of GRW and CSL theories, since it is deterministic (no Wiener
processes are assumed); the random character of a result of measurement is
then just a consequence of the random value of the initial positions. The
dynamics also suppresses the macroscopic \textquotedblleft empty
waves\textquotedblright\ of the dBB theory when they correspond to macroscopic
systems (waves that never play a role in the future), eliminating any
conceptual difficulty concerning the interpetation of these waves. It
therefore seems to provide a simple and a reasonably plausible mechanism for
quantum collapse.

\section{Equations of motion with collapse}

The change of the dynamics of the quantum state we propose is continuous (as
opposed to the standard projection postulate), but nevertheless manages to
suppress the cat paradox and the von Neumann's infinite regress.\ It then
becomes possible to consider, as in the GRW and CSL theories \cite{GRW, CSL},
that the state vector directly represents physical reality.

The study of Bohmian positions actually provides a convenient indicator of
Schr\"{o}dinger's \textquotedblleft ridiculous cases\textquotedblright,\ and
therefore suggests a way to avoid them. Consider the Bohmian positions of the
atoms contained in the glass bottle containing the poison that may kill the
cat.\ After some time, the linear Schr\"{o}dinger equation predicts a
superposition of states where the bottle is broken and intact; the probability
densities of the constituent atoms are spread between different locations in
space.\ By contrast, for each realization of the experiment, the dBB theory
predicts that the bottle is either broken or intact; the Bohmian positions of
its constituents atoms remain bunched together in only one of the possible
locations.\ This means that, in configuration space, one of the components of
the quantum wave function propagates accompanied by Bohmian positions, while
the other propagates \textquotedblleft alone\textquotedblright; it has become
what Bohm calls an \textquotedblleft empty wave\textquotedblright%
\ \cite{Bohm}.\ In order to introduce macroscopic uniqueness in the
propagation of the wave function, we will therefore introduce a dynamical
process that suppresses the components that are propagating too far from the
Bohmian positions; this will force a better match between the evolution of the
wave function and that of the positions.

\subsection{Bohmian localization operator $L$}

We consider a system of $N$ identical particles associated with a quantum
field operator $\Psi\left(  \mathbf{r}\right)  $ defined at each point
$\mathbf{r}$ of ordinary 3D space.\ When the system is in state $\left\vert
\Phi\right\rangle $, the local (number) density $D_{\Phi}\left(
\mathbf{r}\right)  $ of particles at $\mathbf{r}$ is:%
\begin{equation}
D_{\Phi}\left(  \mathbf{r}\right)  =\frac{\left\langle \Phi\right\vert
\Psi^{\dagger}\left(  \mathbf{r}\right)  \Psi\left(  \mathbf{r}\right)
\left\vert \Phi\right\rangle }{\left\langle \Phi\right.  \left\vert
\Phi\right\rangle }\label{sdap-1}%
\end{equation}
In dBB theory, the local density $D_{B}\left(  \mathbf{r}\right)  $ of Bohmian
positions is a sum of delta functions:%
\begin{equation}
D_{B}\left(  \mathbf{r,}t\right)  =\sum_{n=1}^{N}\delta\left(  \mathbf{r}%
-\mathbf{q}_{n}\right)  \label{sdap-2}%
\end{equation}
where the sum runs over all $N$ particles with Bohmian position $\mathbf{q}%
_{n}\left(  t\right)  $. We wish to introduce a dynamics that favors
evolutions where $D_{\Phi}\left(  \mathbf{r}\right)  $\ is attracted towards
regions where $D_{B}\left(  \mathbf{r}\right)  $ is high, with a space average
suppressing the microscopic fluctuations of $D_{B}\left(  \mathbf{r}\right)
$. For this purpose, we introduce an averaging length $a_{L}$ and the
following integral of $D_{B}$:%
\begin{equation}
N_{B}\left(  \mathbf{r,}t\right)  =\int d^{3}r^{\prime}~e^{-\left(
\mathbf{r}-\mathbf{r}^{\prime}\right)  ^{2}/\left(  a_{L}\right)  ^{2}}%
D_{B}\left(  \mathbf{r}^{\prime},t\right)  =\sum_{n=1}^{N}e^{-\left(
\mathbf{r}-\mathbf{q}_{n}\right)  ^{2}/\left(  a_{L}\right)  ^{2}%
}\label{sdap-500}%
\end{equation}
The order of magnitude of $N_{B}\left(  \mathbf{r,}t\right)  $ is the number
of Bohmian positions within a volume $\left(  a_{L}\right)  ^{3}$ around point
$\mathbf{r}$; we have $0\leq$ $N_{B}\left(  \mathbf{r,}t\right)  \leq N$. \ We
then introduce the localization operator $L\left(  t\right)  $ by:%
\begin{equation}
L\left(  t\right)  =\int d^{3}r~N_{B}\left(  \mathbf{r},t\right)
~\Psi^{\dagger}\left(  \mathbf{r}\right)  \Psi\left(  \mathbf{r}\right)
=\sum_{p=1}^{N}N_{B}\left(  \mathbf{R}_{p},t\right)  \label{sdap-500-defn-L}%
\end{equation}
($\mathbf{R}_{p}$ is the position operator of particle $p$). This operator
combines the quantum density operator $\Psi^{\dagger}\left(  \mathbf{r}%
\right)  \Psi\left(  \mathbf{r}\right)  $\ with the classical averaged density
$N_{B}\left(  \mathbf{r},t\right)  $.\ It has the form of a single-particle
potential energy operator; $L\left(  t\right)  $ multiplies any wave function
$\Phi\left(  \mathbf{r}_{1},\mathbf{r}_{2},..,\mathbf{r}_{N}\right)  $ by the
the sum over $p$ of the individual potentials $N_{B}\left(  \mathbf{r}%
_{p},t\right)  $.

The result depends on the relative positions of all Bohmian positions
$\mathbf{q}_{n}$.\ When they are all at large relative distances from each
other (larger than $a_{L}$), the $\mathbf{r}$ dependence of $N_{B}$ exhibits a
series of bumps centered on each $\mathbf{q}_{n}$, each of height unity, and
separated by intervals where $N_{B}$ practically vanishes; if none of the
variables $\mathbf{r}_{p}$ falls inside one of these bumps, the effect of
$L\left(  t\right)  $ is merely to cancel the wave function; if $p$ of them
fall inside these bumps, the effect of $L\left(  t\right)  $ is roughly a
multiplication by $p$.\ When, at the other extreme, all Bohmian positions are
clustered together inside a single volume $\mathcal{V}$ of size smaller than
$\left(  a_{L}\right)  ^{3}$, the $\mathbf{r}$ dependence of $N_{B}$ now
exhibits a single bump of height $N$, and the effect of $L\left(  t\right)  $
is more focussed in space; for instance, in the region of configuration space
where all variables $\mathbf{r}_{p}$ of $\Phi\left(  \mathbf{r}_{1}%
,\mathbf{r}_{2},..,\mathbf{r}_{N}\right)  $ fall inside $\mathcal{V}$, the
wave function is multiplied by $N^{2}$. A more frequent case occurs when all
positions are spread almost uniformly (at a scale $a_{L}$) within a certain
volume $\mathcal{V}$, as is for instance the case if the physical system is a
piece of a solid containing a large number $N_{B}$ of particles within a
volume $\left(  a_{L}\right)  ^{3}$.\ The localization effect then occurs
inside volume $\mathcal{V}$, with a rate that is propostional to $N\times
N_{B}$.\ Again we obtain a rate that is quadratic in the number of particles;
this fast dependence plays an important role in the sudden collapse mechanism
we discuss below.

\subsection{Modified attractive quantum dynamics}

We wish to introduce a dynamics that favors evolutions where $D_{\Phi}\left(
\mathbf{r}\right)  $\ is attracted towards regions where $N_{B}\left(
\mathbf{r}\right)  $ is high.\ For this purpose, we add to the usual
Hamiltonian $H\left(  t\right)  $ a localization term proportional to
$L\left(  t\right)  $ and write the modified Schr\"{o}dinger equation:%
\begin{equation}
i\hslash\frac{d}{dt}\left\vert \Phi\left(  t\right)  \right\rangle =\left[
H\left(  t\right)  +i\hslash\gamma_{L}~L\left(  t\right)  \right]  \left\vert
\Phi\left(  t\right)  \right\rangle \label{sdap-4}%
\end{equation}
where $\gamma_{L}$ is a constant localization rate and $a_{L}$ a localization
length. The new term in the Hamiltonian increases the modulus of the wave
function in regions where the Bohmian density is large. It is not Hermitian,
and no longer conserves the norm of $\left\vert \Phi\right\rangle $.\ We
consider that $\left\vert \Phi\right\rangle $ defines the direction in the
space of states (a one-dimension subspace of this space, what von Neumann
calls a \textquotedblleft ray\textquotedblright), so that its norm is
irrelevant.\ Nevertheless, if desired, one can easily obtain a normalized
state vector $\left\vert \overline{\Phi}\right\rangle $, which obeys the
following equation of evolution:%
\begin{equation}
i\hslash\frac{d}{dt}\left\vert \overline{\Phi}\left(  t\right)  \right\rangle
=\left[  H\left(  t\right)  +\overline{H}_{L}\left(  t\right)  \right]
\left\vert \overline{\Phi}\left(  t\right)  \right\rangle
\label{sdap-equa-conservant-norme}%
\end{equation}
where:%
\begin{equation}
\overline{H}_{L}\left(  t\right)  =i\hslash\gamma_{L}\int d^{3}r~\left[
\Psi^{\dagger}\left(  \mathbf{r}\right)  \Psi\left(  \mathbf{r}\right)
-D_{\Phi}\left(  \mathbf{r}\right)  \right]  N_{B}\left(  \mathbf{r,}t\right)
\label{sdap-5-bis}%
\end{equation}
The only difference with (\ref{sdap-4}) is that the operator $\Psi^{\dagger
}\left(  \mathbf{r}\right)  \Psi\left(  \mathbf{r}\right)  $ has been replaced
by that appearing inside the brackets, which is actually nothing but the
operator associated with the fluctuation of the local density.

Equation (\ref{sdap-4}) is linear but time-dependent, even if the Hamiltonian
$H$ is time independent, since the Bohmian positions and thus $N_{B}\left(
\mathbf{r},t\right)  $ depend on time.\ The norm-conserving version
(\ref{sdap-5-bis}) is non-linear because $D_{\Phi}\left(  \mathbf{r}^{\prime
}\right)  $ depends on the state vector $\left\vert \Phi\right\rangle $. The
modified dynamics we study in this article is defined by these time
differential equations.

\subsection{Coupled evolutions}

The positions $\mathbf{q}_{n}$ evolve according to the usual Bohmian equation
of motion:%
\begin{equation}
\frac{d\mathbf{q}_{n}\left(  t\right)  }{dt}=\hslash\frac{\overrightarrow
{\bigtriangledown}_{n}\xi}{m} \label{sdap-5-ter}%
\end{equation}
where $\xi\left(  \mathbf{r}_{1},\mathbf{r}_{2},..,\mathbf{r}_{N}\right)  $ is
the phase of the wave function $\Phi\left(  \mathbf{r}_{1},\mathbf{r}%
_{2},..,\mathbf{r}_{N}\right)  $. Since $L\left(  t\right)  $is diagonal and
real in the position representation, it does not change the phase of the wave
function, but only its modulus.\ The localization process therefore does not
change the Bohmian velocities directly.\ It nevertheless changes them
indirectly, because the evolution of the phase $\xi\left(  \mathbf{r}\right)
$ depends on the Laplacian of the modulus $\left\vert \Psi\right\vert $ of the
wave function:%
\begin{equation}
\hslash\frac{\partial\xi}{\partial t}+\frac{\hslash^{2}}{2m}\left[  \sum
_{n}\left(  \left(  \overrightarrow{\bigtriangledown}_{n}~\xi\right)
^{2}-\frac{\Delta_{n}\left\vert \Psi\right\vert }{\left\vert \Psi\right\vert
}\right)  +V\right]  \label{sdap-9}%
\end{equation}
where $\overrightarrow{\bigtriangledown}_{n}$ and Laplacian $\Delta_{n}$
contain derivatives with respect to the $3$ coordinates of particle $n$; $V$
is the usual potential operator.\ The term in $\Delta_{n}\left\vert
\Psi\right\vert /\left\vert \Psi\right\vert $ is often called the
\textquotedblleft quantum potential\textquotedblright. The effect of the
localization process is to introduce smooth variations of $\left\vert
\Psi\right\vert $ taking place over a distance of the order of $a_{L}$.\ This
spreads the Fourier components of the wave function over a range $\Delta
k\simeq1/a_{L}$; only particles having de Broglie wavelengths $\lambda$ that
are of the order of (or larger than) $a_{0}$ can undergo an appreciable change
of their Bohmian velocity. With the mesoscopic value (\ref{sdap-6}) chosen for
$a_{L}$, this corresponds to very low velocities; they are transferred to the
Bohmian positions at a rate $\gamma_{L}$, for which we will choose below a
very small value.\ Altogether, after time integration, the localization terms
produces very tiny changes of the Bohmian positions.

A general remark is that, in the limit $a_{L}\rightarrow\infty$, the
localization term has no effect: in (\ref{sdap-500}), $N_{B}\left(
\mathbf{r}\right)  $ then becomes equal to the number of particles $N$ (a
constant) and, in (\ref{sdap-500-defn-L}) $L\left(  t\right)  $ becomes the
products $N\widehat{N}$ (where $~\widehat{N}$ is the operator associated with
the total number of particles).\ The right hand side of (\ref{sdap-5-bis})
then becomes proportional to $N\left(  \widehat{N}-N\right)  $, which gives
zero when acting on any ket $\left\vert \overline{\Phi}\left(  t\right)
\right\rangle $ with a fixed number of particles; nothing is then changed with
respect to standard Schr\"{o}dinger dynamics.

\section{Collapse in small or large systems}

We now assume investigate the dynamics of physical systems obeying the
modified Schr\"{o}dinger equation (\ref{sdap-4}).\ As in GRW and CSL theories,
our purpose is to check that it is possible to find plausible values of the
two parameters $\gamma_{L}$ and $a_{L}$; by this we mean values for which no
contradiction occurs with the enormous body of experimental data agreeing with
quantum mechanics (sometimes with an incredible precision of $10^{-12}$!).
What is needed is a compromise between opposite requirements: a localization
dynamics that has fantastically small effects on microscopic systems, but
nevertheless produces a sufficiently fast collapse of superpositions of
macroscopically different states. We will choose values inspired by those
often chosen in GRW and CSL theories, namely:\
\begin{align}
\gamma_{L}  &  =10^{-16}~s^{-1}\nonumber\\
a_{L}  &  \simeq10^{-6}~m \label{sdap-6}%
\end{align}
Our purpose here is not to define accurate values of these constants; we just
wish to show that there is a wide range of values that are compatible with the
above criteria.

\subsection{Microscopic system}

Consider first a microscopic system, atom, molecule or nucleus, with a wave
function of all the constituent particles extending over a range $a_{0}\ll
a_{L}$. Since the $\mathbf{q}_{n}$'s can never reach regions of space where
the wave function vanishes, they also remain localized in a region of space of
dimension $a_{0}$.\ This corresponds to the case mentioned above, where
$N_{B}\left(  \mathbf{r}\right)  $ is of the order of the total number of
particles $N$ in a domain of size $a_{L}$ centered on the atom, and tends
rapidly to zero outside of this domain.\ In the limit $a_{0}/a_{L}%
\rightarrow0$, we have seen that $L\left(  t\right)  \rightarrow N\widehat{N}%
$, so that in (\ref{sdap-4}) the localization term has no effect on the wave
function (except a multiplication by an overall factor without any physical
consequence).\ If $a_{0}/a_{L}\ll1$, the exponential in (\ref{sdap-500}) can
be approximated by $1-c\left(  a_{0}/a_{L}\right)  ^{2}$, where the term in
$1$ does not contribute (this is the limit $a_{L}\rightarrow\infty$), and
where $c\simeq1$ (the exact value of $c$ depends on the Bohmian
positions).\ So, retaining only the term in $\left(  a_{0}/a_{L}\right)  ^{2}%
$, we see that the parts of the wave function at the periphery of the atom are
reduced at a rate $\gamma$ given by:%
\begin{equation}
\gamma\lesssim\gamma_{L}\left(  \frac{a_{0}}{a_{L}}\right)  ^{2}N^{2}
\label{sdap-7}%
\end{equation}
while the parts near the center of the atom remain unaffected.

For a small atom (Hydrogen or Helium for instance), with values (\ref{sdap-6}%
), $a_{0}/a_{L}\simeq10^{-4}$ so that $\gamma\leq10^{-24}N^{2}$, where $N$ is
a few units; this rate is clearly extremely low and undetectable.\ For a
molecule, a size of $10$ nm is already large, which corresponds to
$a_{0}/a_{L}\simeq10^{-2}$ and to $\gamma\leq10^{-20}N^{2}$; even with a
number of constituents (protons, neutrons) of the order of $10^{4}$, we still
obtain an extremely small rate.

Now consider an interference experiment made with the same microscopic
system.\ In the interferometer, its wave function is localized at the same
time in very different regions of space; in one of these regions,
$N_{B}\left(  \mathbf{r}\right)  $ is equal to $N$ as above, but in the other
it is zero. This clearly introduces an imbalance between the full wave, which
increases at a rate $\gamma_{L}N^{2}$, and the (constant) empty wave.\ The
rate of growth of this imbalance is:%
\begin{equation}
\gamma\simeq\gamma_{L}~N^{2} \label{sdap-7-2}%
\end{equation}
Therefore, even for a long experiment lasting one second, if $N<10^{7}$, the
localization rate remains negligible, and the interference takes place as in
standard Schr\"{o}dinger dynamics; but, for larger values of $N$, this
dynamics predicts that the contrast of fringes should decrease and vanish in
the limit $N\gg1/\sqrt{\gamma_{L}t}$.

\subsection{Macroscopic system}

\label{SDAP-macroscopic-system}The orders of magnitude are completely
different for macroscopic systems.\ Consider for instance the pointer of a
measurement apparatus; after measurement it may reach (for instance) two
different positions that are $10$ microns apart from each other. The solution
of the linear Schr\"{o}dinger equation has components where the particles of
the pointer are, either in one region of space, or in another; big
fluctuations of the local density of particles then take place.\ By contrast,
in a given realization of the experiment, the corresponding Bohmian variables
remain all clustered in the vicinity of only one of these two positions.\ They
necessarily remain together because the state vector has no components where
some of the pointer particles are in one site, some in the other: this is
forbidden by the cohesion forces inside the material forming the pointer.\ So,
when the measurement is performed, one component of the wave function resides
in the same region of configuration space as many Bohmian variables (this is a
\textquotedblleft full wave\textquotedblright), but the other component in
region where the density of Bohmian positions is zero (this is an empty wave).

For the \textquotedblleft full wave\textquotedblright, in the localization
term in the right hand side of (\ref{sdap-4}), the relevant values of
$\mathbf{r}$ in the integral are those in the region of space where this wave
propagates.\ Since $N_{B}$ has significant values in this region, this term
increases the modulus of the full wave.\ We have seen that $L\left(  t\right)
$ is a potential operator, the sum of $N$ individual potentials that are equal
to $N_{B}$; its effect on the wave function is to multiply it by the product
$N_{B}N_{P}$, where $N_{P}$ is of the order of the number of particles in the
pointer and $N_{B}$,\ the number of its particles in volume $\left(
a_{L}\right)  ^{3}$. As for the \textquotedblleft empty wave\textquotedblright%
, the relevant values of $\mathbf{r}$ are those in the region of space where
$N_{B}$ is zero, and the localization term \ does not have any effect.\ When
the measurement result is fully registered in the position of the particles of
the pointer, the relative weight of the full wave is therefore increased
exponentially with a time rate $\gamma$ of the order of:%
\begin{equation}
\gamma\simeq\gamma_{L}\ N_{B}~N_{P} \label{sdap-8}%
\end{equation}
If we choose conservatively small values $N_{P}=10^{20}$, $N_{B}=10^{11}$, we
obtain a very fast rate $\gamma\simeq10^{15}$; the dynamical equation leads to
an extremely fast collapse of the wave function!

After collapse, the state vector of the physical system continues to evolve
under the influence of the Hamiltonian $H\left(  t\right)  $ and of the
localization term.\ Their effects are very different, since the full
Hamiltonian $H\left(  t\right)  $ contains interactions between the particles
and directly controls the correlations between the particles, while the
localization term is just a single particle operator, similar to mean field
operator (with an anti-Hermitian contribution to the evolution).\ In a low
compressibility solid or liquid, the average energy $\left\langle
H\right\rangle $ varies very rapidly as a function of the average distance
between the particles; the weak localization term cannot change $\left\langle
H\right\rangle $ significantly, so that the quantum average distance between
the particles remains practically constant, leading to a density $D_{\Phi
}\left(  \mathbf{r}\right)  $ that is almost uniform in the volume of the solid.

We now examine how the ensemble of $\mathbf{q}_{n}$'s reacts to this wave
function in configuration space.\ We have mentioned above that the collapse
process may affect the $\mathbf{q}_{n}$'s indirectly.\ Nevertheless, Towler,
Russell and Valentini \cite{Valentini} have shown that a fast relaxation
process tends to constantly bring back any distribution of the $\mathbf{q}%
_{n}$'s in configuration space towards that of quantum
equilibrium\footnote{These authors study the evolution of the distribution of
the $\mathbf{q}_{n}$'s when the wave function obeys the standard
Schr\"{o}dinger equation (without any localization term).\ Since they predict
relatively short relaxation times, their conclusions for microscopic objects
should not be changed significantly by the introduction of a localization term
having a very small coupling constant.} (the coarse graining considered in
this reference is immediately provided here by the average over the
localization length $a_{L}$); the distribution of the $\mathbf{q}_{n}$'s then
closely follows the square of the wave function.\ As a consequence, in
ordinary space, a spatially uniform $D_{\Phi}\left(  \mathbf{r}\right)  $
results in a practically uniform distribution of $N_{B}\left(  \mathbf{r}%
\right)  $. The only effect of the term in $L\left(  t\right)  $ is then to
localize all particles within the volume of the whole solid -- but this is
nothing but what is usually done in many-body physics, when one assumes that
the physical system is contained in a box.\ The standard Schr\"{o}dinger
dynamics therefore applies with no change.

\subsection{Bose-Einstein condensate}

Consider now a gaseous Bose-Einstein condensate that is partly reflected by
Bragg scattering on a laser standing wave \cite{Philips}. A matter wave is
then split into two coherent parts, which can propagate at macroscopic
distances and interfere again if recombined.\ We must check that the
localization term does not destroy the coherence, which would be in
contradiction with the experimental observations. The major difference with
the preceding case is that the atoms in the condensate propagate almost
freely, and that no process forces all of them to go in the same direction;
there is no reason to find all atoms in the same output beam.\ The
distribution of Bohmian positions then closely follow the quantum distribution
(a Poisson distribution of populations in the two output atomic clouds).\ In
other words, none of the two waves becomes empty; each of them travels
accompanied by a Bohmian density that is proportional to its intensity.
Moreover, the number of atoms involved in these experiments is of the order of
$10^{5}$, in a volume that is comparable with $a_{L}$.\ So, even in the
absence of Bohmian density, relation (\ref{sdap-8}) would lead to
$\gamma\simeq10^{-6}$, still a very small rate.\ The collapse predicted in the
previous case does not take place here, and the two waves propagate as
coherent classical waves.

\subsection{Measurement, Born rule}

A soon as a microscopic quantum system $S$ interacts with a measurement
apparatus $M$, it becomes entangled with some of its particles: each
measurement eigenstate $\left\vert s_{i}\right\rangle $ of $S$ becomes
associated with different states of $M$. As we have seen, the usual
Schr\"{o}dinger dynamics is not affected until a large number $N_{m}$ of
particles of $M$ is involved in this process.\ But, since the very purpose of
a measurement apparatus is to transfer information to a macroscopic scale,
entanglement progresses rapidly within the measurement apparatus (it
propagates by \textquotedblleft contagion\textquotedblright\ between mutually
interacting neighbor particles, as discussed in \cite{Omnes}); it quickly
reaches a mesoscopic and macroscopic scale.\ In standard quantum mechanics,
this phenomenon occurs in parallel in all \textquotedblleft measurement
channels\textquotedblright\ (values of the index $i$); various branches of the
state vector develop, each associated with one of the measurement eigenstates
$\left\vert s_{i}\right\rangle $. In Bohmian mechanics, this propagation
induces a motion of the Bohmian positions; nevertheless, for each realization
of the experiment, the motion occurs only within one measurement channel (all
the other channels correspond to empty waves having no effect on the Bohmian
positions).\ The changes of the values of $N_{B}\left(  \mathbf{r,}t\right)  $
are therefore those associated with a a single branch of the state vector.\ In
other words, for each realization of the measurement, the time dependence of
$N_{B}\left(  \mathbf{r,}t\right)  $ is different and, in (\ref{sdap-4}), the
operator $L\left(  t\right)  $ tends to localize the wave function in a
different way.

Consider now a time when the values of $N_{B}\left(  \mathbf{r,}t\right)  $
have become significantly different for each value of $i$ in macroscopic
regions of space; by the process discussed above for macroscopic systems, the
effect of the localization process on the wave function is then very fast, and
selects in the wave function the component associated with a single value of
$i$ (one result of measurement). State vector reduction has then taken place.
As a whole, the localization process is similar to fast freezing of a liquid:
the suppression of the \textquotedblleft empty\textquotedblright\ components
of the wave function occurs with a time constant that is initially completely
negligible, but grows faster than linearly in time (exponentially at the
beginning of propagation of entanglement).\ 

Now, instead of a single realization of the experiment, consider an ensemble
of realizations. When the initial Bohmian positions of the particles are
chosen randomly according to the quantum distribution, the dBB theory exactly
reproduces the Born rule of standard quantum mechanics.\ As a consequence,
before the sudden localization process of the wave function takes place, the
effects of the localization term on the positions are still negligible, and
the standard quantum probability gives the proportion of realizations where
the $N_{m}$ Bohmian positions of the measurement apparatus reach the region of
space associated with a measurement eigenstate $\left\vert s_{i}\right\rangle
$.\ For the wave function, this region determines the branch that will be
enhanced by the localization process, while all the empty waves are
cancelled.\ As a consequence, the collapse of the wave function takes place
randomly towards one of the regions of space associated with the various
possible results of measurement, with a probability given by the standard Born
rule.\ Without conflict with the predictions of quantum mechanics, we can then
consider the field associated with $\left\vert \overline{\Phi}\left(
t\right)  \right\rangle $ (within a phase factor) as directly related to
physical reality (in configuration space).

\subsection{Effects of decoherence, stabilization}

Decoherence plays a very important role in standard quantum mechanics
\cite{Zurek}, even if it does not ensure macroscopic uniqueness.\ The dynamics
resulting from (\ref{sdap-4}) provides this uniqueness: even if no measurement
is performed, the interactions of the physical system with its environment
produce collapse through the localization mechanism.\ To illustrate this with
an example, let us consider a system of $3$ particles, two with Bohmian
positions at a small relative distance (smaller than $a_{L}$; we can then
consider that these positions coincide: $\mathbf{q}_{1}=\mathbf{q}%
_{2}=\mathbf{u}$), and a third far away ($\mathbf{q}_{3}=\mathbf{v}%
\neq\mathbf{u}$). We assume that $\left\vert \varphi_{u}^{0}\right\rangle $ is
an individual state localized around $\mathbf{q}_{1}=\mathbf{q}_{2}$ and that
$\left\vert \varphi_{v}\right\rangle $ is another individual state localized
around $\mathbf{q}_{3}$. The initial quantum state of the $3$ particles is a
superposition:%
\begin{align}
\left\vert \Phi\left(  0\right)  \right\rangle  &  =\alpha\left\vert
1:\varphi_{u}^{0}\right\rangle \left\vert 2:\varphi_{u}^{0}\right\rangle
\left\vert 3:\varphi_{u}^{0}\right\rangle \left\vert E_{u}^{0}\right\rangle
\left\vert E_{v}^{0}\right\rangle \nonumber\\
&  \ \ \ \ \ \ \ \ \ +\beta\left\vert 1:\varphi_{u}^{0}\right\rangle
\left\vert 2:\varphi_{u}^{0}\right\rangle \left\vert 3:\varphi_{v}%
^{0}\right\rangle \left\vert E_{u}^{0}\right\rangle \left\vert E_{v}%
^{0}\right\rangle \label{sdap-superposition}%
\end{align}
where $\alpha$ and $\beta$ are arbitrary complex coefficients; the kets
$\left\vert E_{u}^{0}\right\rangle $ and $\left\vert E_{v}^{0}\right\rangle $
denote the initial states of the environments surrounding the two regions of
space of $u$ and $v$. Note that the first component of the superposition
contains a mismatch between the localization of the particles in the state
vector (the three at the same site) and their Bohmian positions (only two
particles have the same Bohmian position). If the system remained totally
isolated, this mismatch could continue to exist forever, especially since the
first component is favored by a larger local value of $N_{B}$.

But this does not happen in practice, since a physical system always interacts
with its environment.\ When the particles become locally entangled with their
environment, the state vector $\left\vert \Phi\left(  t\right)  \right\rangle
$ becomes:%
\begin{align}
\left\vert \Phi\left(  t\right)  \right\rangle  &  =\sum_{i,j,k}\alpha
_{i,j,k}^{\prime}\left\vert \vphantom{E_{u}^{i,j,k}}1:\varphi_{u}%
^{i}\right\rangle \left\vert \vphantom{E_{u}^{i,j,k}}2:\varphi_{u}%
^{j}\right\rangle \left\vert 3:\varphi_{u}^{k}\right\rangle \left\vert
E_{u}^{i,j,k}\right\rangle \left\vert E_{v}^{0}\right\rangle \nonumber\\
&  +\sum_{l,m,n}\beta_{l,m,n}^{\prime}\left\vert 1:\varphi_{u}^{l}%
\right\rangle \left\vert \vphantom{E_{u}^{i,j,k}}2:\varphi_{u}^{m}%
\right\rangle \left\vert \vphantom{E_{u}^{i,j,k}}3:\varphi_{v}^{n}%
\right\rangle \left\vert E_{u}^{l,m,n}\right\rangle \left\vert E_{v}%
^{_{l,m,n}}\right\rangle \label{sdap-superposition-a-t}%
\end{align}
where the $\alpha_{i,j,k}^{\prime}$ and $\beta_{l,m,n}^{\prime}$ are the
coefficients characterizing these local entanglements; the states where the
upper index $0$ has been replaced by other indices denote the states taking
part in this entanglement.\ In the first line, the state $\left\vert E_{v}%
^{0}\right\rangle $ of the environment at $v$ remains unchanged, since for
this component the wave function contains no particle in this site; as for the
state $\left\vert E_{u}^{0}\right\rangle $ of the environment around $u$, it
is changed in different ways, depending if this environment interacts with $2$
or $3$ particles.\ Now, the Bohmian velocities are obtained from the phase of
the wave function at the point of configuration space defined by all Bohmian
positions; since we have assumed that one particle is in site $v$, only the
phase of the components in the second line are relevant to this calculation
(all components of the first line are empty waves). As a consequence, while
the states of the environment in the second line propagate in space
accompanied by Bohmian positions, the states in the first line do not.\ While
these positions move in the environment, the localization term then favors the
components of the second line, exponentially reducing the relative weight of
those in the first line. If the environment is macroscopic, the corresponding
collapse is very fast; in practice, only the components where the positions of
particles are close to the Bohmian positions survive.\ Through this process,
the environment has a stabilizing effect of the wave function towards the
Bohmian trajectories of all particles; it rapidly corrects any mismatch
between the propagation of the state vector and that of Bohmian positions.

\section{Discussion}

The localization process contained in equation (\ref{sdap-4}) occurs in
ordinary 3D space, as in CSL\ theory, but is also very different.
CSL\ introduces the simultaneous action of an infinite number of random
processes modifying the wave function in all points of space, and described by
Wiener processes.\ A \textquotedblleft probability rule\textquotedblright%
\ ensures that the processes cooperate in space in order to conserve the
maximum possible norm to the state vector, which leads to the usual Born
rule.\ Gisin \cite{Gisin} also assumes the presence of an additional random
term in the Schr\"{o}dinger equation; the evolution of the state vector
depends on a stochastic Wiener process which, conversely, has an evolution
that depends on the state vector.

Here, no stochastic process is introduced, and the evolution is entirely
deterministic.\ Randomness arises only from the initial distribution of the
dBB positions. The time dependence of the localization term equation arises
from the time dependence of the Bohmian density $N_{B}\left(  \mathbf{r,}%
t\right)  $, which is different in different realizations of the same
experiment.\ If no important fluctuation of the Bohmian density occurs from
one realization to another, the localization term remains completely
negligible; but, in measurement-like situations, is quickly collapses the wave function.

Another difference with GRW\ and CSL\ theories is that the localization
process does not take place with a constant localization length.\ It occurs
over a variable range that depends on the space distribution of matter; for
instance, no localization happens anymore when the density of the system
cecomes perfectly uniform.\ In a solid, as we discussed at the end of
\S \ \ref{SDAP-macroscopic-system}, the localization takes place inside a
volume that is nothing but the whole volume of the solid. As a consequence,
the effects of the added term in the evolution can be much softer than in
standard localization theories.\ It corresponds to a much smaller spontaneous
heating of material bodies than GRW\ and CSL -- but it nevertheless predicts
the disappearance of interferences for macroscopic objects roughly in the same
situations as these theories. Another difference is that the stochastic
processes of GRW\ and CSL are Markovian processes, with no memory; here, the
Bohmian variables introduce some memory, since their positions depend on
previous velocities, and therefore of previous values of the wave
function.\ As a consequence, the simple Lindblad forms for the evolution of
the density operator in GRW\ and CSL theories are no longer obtained.

The equations obey Galilean relativity (absolute time), which is logically
consistent for a modification of the Schr\"{o}dinger equation; it would
nevertheless be interesting to extend them to Einsteinian relativity, for
instance with a time delayed density localization term.

Allori et al. \cite{Allori} have proposed an interesting related idea:
multiply the usual wave function in configuration space by a function centered
around all individual Bohmian positions; this truncation provides a
\textquotedblleft collapsed\textquotedblright\ wave function.\ The two wave
functions are directly related at any time $t$; they do not have really
independent evolutions.\ With this dynamics, the collapsed waves that have
been truncated may reappear later to create interference; the new wave
function is no longer symmetrical by exchange of identical particles. This
scheme depends only on one parameter (the localization length $a_{L}$) instead
of two in our case; also, the collapse is performed in configuration instead
of real space. Bedingham \cite{Bedingham-2011} has also combined Bohmian
mechanics with CSL by introducing a stochastic term into the Schr\"{o}dinger
equation that depends on the positions of the particles, as well as non-linear
filtering techniques. Tumulka \cite{Tumulka-2012} has proposed a simpler
version of a stochastic theory that is basically the same.

\textbf{Conclusion}: The equations of evolution proposed here imply that, in
practice, no significant collapse occurs unless a macroscopic fluctuation of
density exists in ordinary space, meaning that entanglement must have
propagated from a microscopic to a macroscopic level.\ This is precisely, by
construction, what a measurement apparatus is supposed to do with a
microscopic system.\ Collapse takes place when quantum macroscopic density
fluctuations occur within the pointer of a measurement apparatuses, but
neither measurements nor observers play any special role.

Different attitudes are possible to interpret such dynamical equations.\ One
is to consider that the Bohmian positions are the essence of reality, as in
the dBB theory.\ The advantage of the proposed dynamics is then to get rid of
the macroscopic \textquotedblleft empty waves\textquotedblright, which persist
forever in the dBB\ theory, while they are supposed to play no physical role
whatsoever; their existence complicates the discussion of the physical reality
of the waves.\ Another attitude is to consider that the directly observable
component of physical reality is the field (the wave function); the other
component of reality would then be the attractive density $N_{B}\left(
\mathbf{r}\right)  $, acting on the field but not directly observable (a sort
of \textquotedblleft dark density\textquotedblright). Bohmian positions may
then be considered as the tool to generate the propagation in real space of an
attractive density $N_{B}\left(  \mathbf{r}\right)  $; this is slightly
reminiscent of de Broglie's ideas on singularities associated with the wave
function and propagating with it (theory of the double solution). The
propagation in space of the field is then free of the counter-intuitive
aspects attached to Bohmian positions (changing direction in free space for
instance).\ Nevertheless, in this view, a naive realism is not restored: the
field remains very different from a classical field, since it does not
propagate in ordinary 3D\ space.

The main purpose of the present work can be seen as a proof of existence: as
GRW and CSL theories have already shown, it is possible to build a simple
theory where waves represent physical reality, while remaining compatible with
present experimental data.\ Our contribution is to show that introducing a
stochastic dynamics is not a necessary condition.\ Macrorealism \cite{Leggett}%
\ can indeed emerge from the dynamics, without elaborate mathematics in the
equations.\ Needless to say, it remains perfectly legitimate to invoke
esthetical or philosophical reasons to maintain the Schr\"{o}dinger dynamics
unchanged, and adhere to one of the interpretations that are available.\ But
one can also prefer to change the dynamics to obtain a completely unified dynamics.

\textbf{Acknowledgments}: the author is very grateful to Philip Pearle for his
expert guidance and very useful advice. LKB is \textquotedblleft Laboratoire
associ\'{e} \`{a} l'ENS, au CNRS, \`{a} l'UPMC et au Coll\`{e}ge de
France\textquotedblright.

\end{document}